\documentclass[conference]{IEEEtran}
\usepackage{subcaption}
\IEEEoverridecommandlockouts
\usepackage{cite}
\usepackage{amsmath,amssymb,amsfonts}
\usepackage{algorithmic}
\usepackage{graphicx}
\usepackage{textcomp}
\usepackage{pgfplots}
\usepackage{xcolor}

\usepackage[bookmarks=false, pdfencoding=auto]{hyperref}

\usepackage[T1]{fontenc}
\def\BibTeX{{\rm B\kern-.05em{\sc i\kern-.025em b}\kern-.08em
    T\kern-.1667em\lower.7ex\hbox{E}\kern-.125emX}}
\pgfplotsset{compat=1.18}

\usepackage{fancyhdr}

\fancypagestyle{firstpage}{

    \fancyhf{} 

    \lhead{\small Published as a conference workshop paper at PerCom 2026} 

}
\begin{document}
\setlength{\columnsep}{0.201in}

\title{Human Presence Detection via Wi-Fi Range-Filtered Doppler Spectrum on Commodity Laptops\\

\thanks{Part of this work received funding from the European Commission Horizon Europe SNS JU projects 6G-SENSES (GA 101139282) and MultiX (GA 101192521).}
}

\author{\IEEEauthorblockN{Jessica Sanson}
\IEEEauthorblockA{\textit{Intel Deutschland GmbH} \\
Munich, Germany \\
jessica.sanson@intel.com}
\and
\IEEEauthorblockN{Rahul C. Shah}
\IEEEauthorblockA{\textit{Intel Corp.} \\
Santa Clara, CA, USA \\
rahul.c.shah@intel.com}
\and
\IEEEauthorblockN{Valerio Frascolla}
\IEEEauthorblockA{\textit{Intel Deutschland GmbH} \\
Munich, Germany \\
valerio.frascolla@intel.com}
}

\maketitle
\thispagestyle{firstpage}
\begin{abstract}

Human Presence Detection (HPD) is key to enable intelligent power management and security features in everyday devices. In this paper we propose the first HPD solution that leverages monostatic Wi-Fi sensing and detects user position using only the built-in Wi-Fi hardware of a device, with no need for external devices, access points, or additional sensors. In contrast, existing HPD solutions for laptops require external dedicated sensors which add cost and complexity, or rely on camera-based approaches that introduce significant privacy concerns. We herewith introduce the Range-Filtered Doppler Spectrum (RF-DS), a novel Wi-Fi sensing technique for presence estimation that enables both range-selective and temporally windowed detection of user presence. By applying targeted range-area filtering in the Channel Impulse Response (CIR) domain before Doppler analysis, our method focuses processing on task-relevant spatial zones, significantly reducing computational complexity. In addition, the use of temporal windows in the spectrum domain provides greater estimator stability compared to conventional 2D Range-Doppler detectors. Furthermore, we propose an adaptive multi-rate processing framework that dynamically adjusts Channel State Information (CSI) sampling rates-operating at low frame rates (10Hz) during idle periods and  high rates (100Hz) only when motion is detected. To our knowledge, this is the first low-complexity solution for occupancy detection using monostatic Wi-Fi sensing on a built-in Wi-Fi network interface controller (NIC) of a commercial off-the-shelf laptop that requires no external network infrastructure or specialized sensors. Our solution can scale across different environments and devices without calibration or retraining. 
\end{abstract}


\section{Introduction}

Personal computing devices, particularly laptops, are increasingly expected to provide intelligent and context-aware functionalities that enhance user experience while optimizing power consumption. A key capability enabling such intelligence is Human Presence Detection (HPD)—the ability to sense whether a user is present, approaching, or departing from a device. Effective HPD enables several use cases\cite{DellExpressSignin2019}:

\begin{itemize}
    \item Wake-on-Approach: Move from sleep or standby to active state as a user approaches a device \cite{WindowsHPS2023}.
    \item Secure Lock-on-Leave: Detect when a user walks away from a laptop and automatically lock the screen to prevent unauthorized access, addressing both security and privacy concerns without no explicit user action\cite{WindowsHPS2023}.
    \item Adaptive Power Management and Enhanced Battery Lifetime: Dynamically adjust system power states based on user presence—reducing display brightness, throttling Central Processing Unit (CPU) performance, or entering sleep when the user is absent\cite{DellExpressSignin2019}.
    \item Privacy-Aware Computing: Enable presence-based features without relying on camera-based monitoring, addressing growing privacy concerns in personal and professional environments.
   
\end{itemize}
Despite the clear benefits of HPD, deploying practical solutions on commodity laptops faces significant technical challenges. Current approaches generally fall into the following three main categories, each with fundamental limitations.

\textbf{Dedicated Hardware (HW) Sensors:}
Commercial laptops usually incorporate specialized HPD sensors, such as Time-of-Flight and infrared modules, to achieve accurate, low-latency proximity detection. While these sensors provide reliable results, their integration increases design complexity and is typically reserved for premium models. Additionally, their sensing direction has a limited field of view, and the dedicated HW further increases power consumption\cite{FTMSense}.

\textbf{Camera-Based Solutions:}
Although this approach utilizes existing HW, it introduces substantial privacy concerns, as users are often uncomfortable with always-on video monitoring. Moreover, these methods are computationally demanding, not well-suited for battery-powered always-on operation\cite{EmadUdDin2023}, struggle in low-light and fail when the camera is covered.

\textbf{Traditional Wi-Fi Sensing:} Wi-Fi Channel State Information (CSI) has emerged as a promising sensing modality for HPD \cite{b0,b4, EmadUdDin2023,FTMSense,Zhu2023Experience,Yang2022EfficientFi,Yang2020DeviceFree}. However while simple statistical approaches (e.g., amplitude variance or phase standard deviation) impose little computational burden, they cannot provide spatial information (range), thus cannot distinguish movement in a relevant desk‐zone vs. irrelevant ambient motion. Hence they often result in high false‑positive rates and also lack velocity cues (approach vs. leaving)\cite{b1,b5,b6,b7,b3}. More sophisticated WiFi‐sensing systems adopt machine‑learning pipelines involving high‑dimensional feature extraction, e.g., Principal Component Analysis or time–frequency transforms, and complex classifiers, e.g., Convolutional Neural Networks, Recursive Neural Networks, or Transformers\cite{Zhu2023Experience,Yang2022EfficientFi,Yang2020DeviceFree,SenseFi,ST_pred}; while these approaches can achieve high detection accuracy in controlled environments, they also impose a heavy computational burden, unsuitable for continuous real-time operation on laptop CPUs. Furthermore, these models generally require extensive environment-specific training data to achieve robustness, thus limiting practical deployment. Additionally, the proposed state of the art solutions for commercial devices are based on bistatic sensing architectures\cite{b1,b5,b6,b7,b3,Zhu2023Experience,Yang2022EfficientFi,Yang2020DeviceFree,SenseFi}, which require the presence of an external access point or router for signal transmission and impose further performance variability due to differences in the environment and device layouts.

Commercial Wi-Fi network interface controller (NICs) can now achieve monostatic radar capabilities for precise localization~\cite{sansonRange,MultiX,6G-SENSES}. However, there has been little evaluation of how this additional range information and active sensing can be applied to HPD. Moreover, standard Range-Doppler Map (RDM) processing requires continuous 2D Fast Fourier Transforms (FFTs) over multiple CSI frames, resulting in high power consumption, unsuitable for always-on laptops.


\begin{figure}[t]
    \centering
    \begin{subfigure}[b]{0.48\linewidth}
        \centering
        \includegraphics[width=\linewidth]{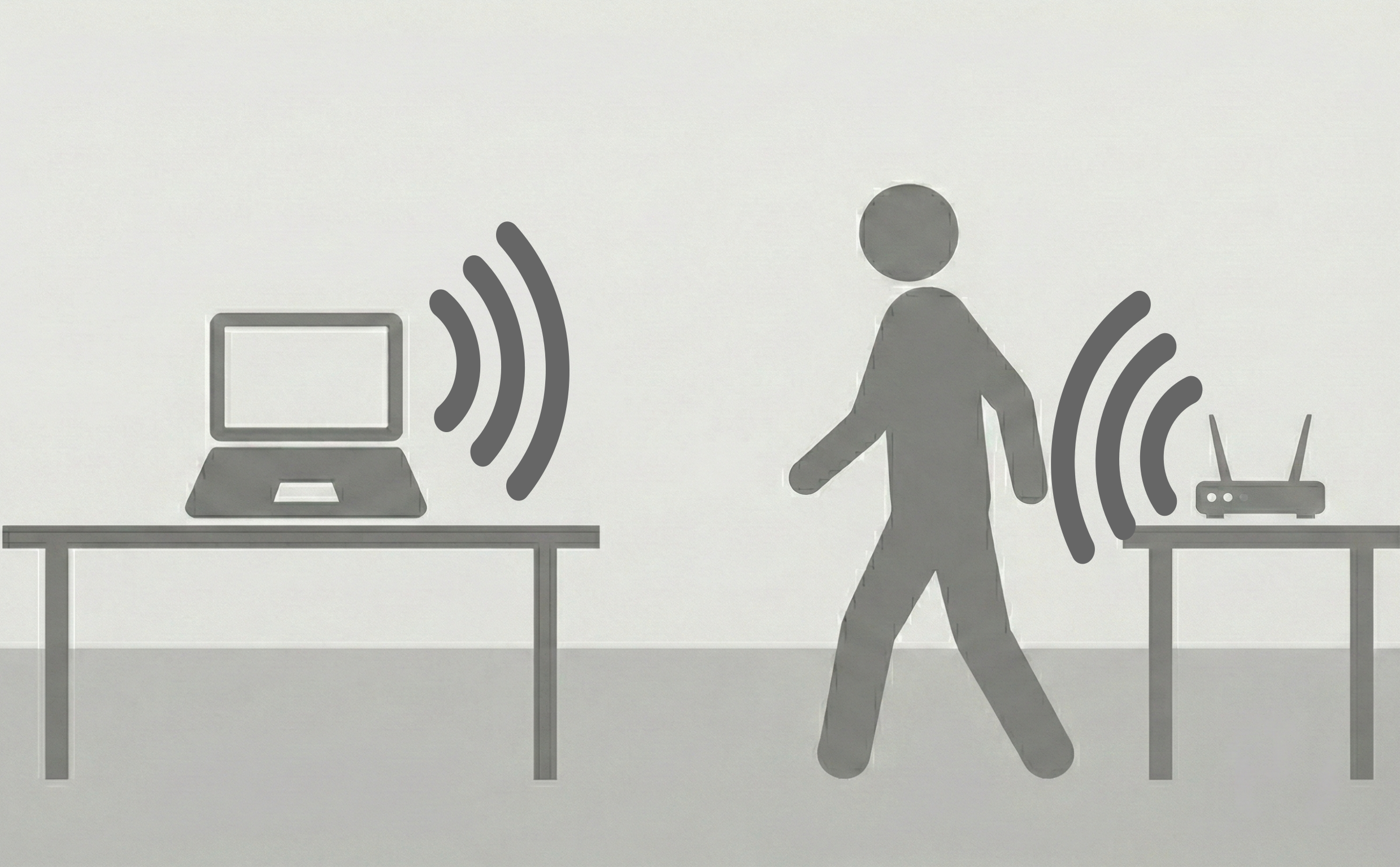} 
        \caption{Bistatic sensing.}
        \label{fig:bistatic_block}
    \end{subfigure}
    \hfill
    \begin{subfigure}[b]{0.49\linewidth}
        \centering
        \includegraphics[width=\linewidth]{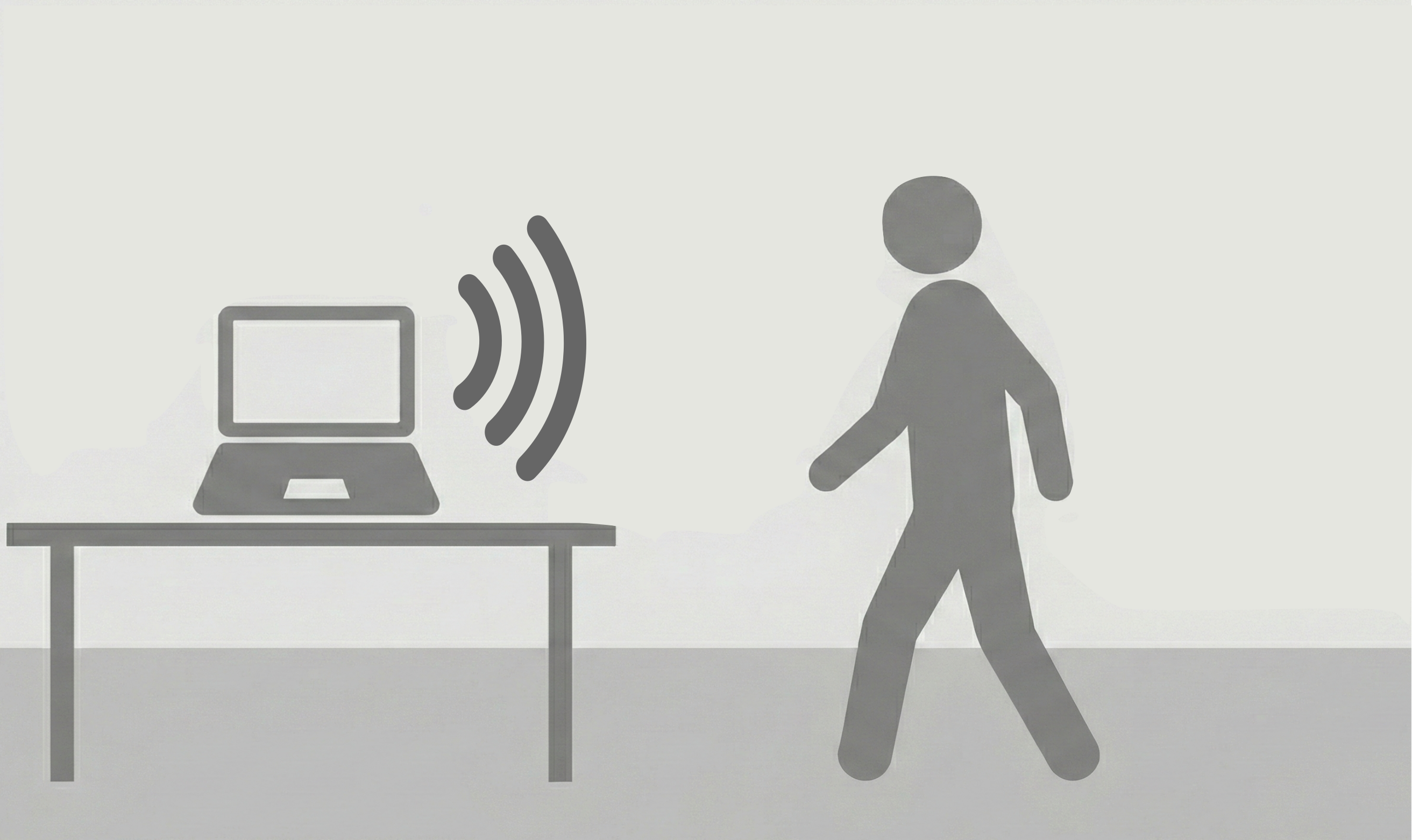} 
        \caption{Monostatic sensing.}
        \label{fig:monostatic_block}
    \end{subfigure}
    \caption{Comparison of Wi-Fi sensing systems.}
    \label{fig:architectures}
\end{figure}


\subsection{Our contribution}

In this paper, we address the gaps described above by implementing HPD on commercial laptops using Wi-Fi-based Range and Doppler estimation. Unlike traditional bistatic systems, our approach (Figure \ref{fig:architectures}) operates on a single laptop sharing the Local Oscillator and baseband to capture both Range and Doppler information (monostatic). We address the efficiency challenge by introducing Range-Filter Doppler-Spectrum (RF-DS), which achieves the advantages of range-selective Doppler estimation with substantially reduced computational overhead by applying range area filtering prior to a simplified 1-D Doppler transform. This approach processes the immediate sensing region in front of the laptop (e.g., 0 to 2 m), effectively ignoring far-field clutter. Furthermore, by leveraging spectrum (time–Doppler) analysis instead of the traditional Range-Doppler Map (RDM), RF-DS enables temporal window estimation, which improves detection stability. The proposed system also incorporates an Adaptive Multi-Frame Rate Framework: it operates in an ultra-low-power mode during idle periods and increases the Wi-Fi frame rate only when motion is detected, supporting reliable user approach and departure detection with minimal battery impact.

The principal contributions of this work are outlined below:
\begin{enumerate}
\item \textbf{First standalone Wi-Fi-based HPD solution:} We demonstrate a novel device-free HPD approach that relies solely on the internal, commercial Wi-Fi NIC of a laptop operating in a monostatic configuration. This enables robust motion and position estimation for user presence detection, making generalized deployment feasible across different environments without the need for any external infrastructure.
\item \textbf{Novel Range-Filtered Doppler Spectrum:} We introduce a  low-complexity method that applies range area filtering and performs range-specific time-Doppler analysis without the need for continuous RDM computation.
\item \textbf{Adaptive Multi-Rate Low Power Framework:} Our architecture dynamically adjusts Wi-Fi sensing rates based on user state, conserving power during idle while rapidly detecting approaching users (from over 6 m).

\end{enumerate}
The CSI measurements used for our evaluation have been made publicly available in our dataset \cite{sanson2026dataset}.

\section{System Model}
\label{sec:background}

\subsection{Wi-Fi Monostatic Sensing}
 Monostatic sensing introduces a radar-like operational mode that enables both range and Doppler estimation, in addition to conventional channel monitoring. Monostatic sensing in Wi-Fi operates with one antenna as the Transmitter (Tx) and the other as the Receiver (Rx) on the same device~\cite{sansonRange}.  The frequency-domain channel matrix of the Long Training Field (LTF) signal, with $N$ subcarriers and $M$ orthogonal frequency-division multiplexing (OFDM) frames reflected by $K$ targets, at subcarrier $n$ and frame $m$, is defined as \cite{sanson2020ofdm}:
\begin{equation}
D(m,n) = \sum_{k=1}^{K} e^{j2\pi T m f_{D,k}} e^{-j2\pi n \Delta f \tau_k} + \tilde{\eta}
\end{equation}
where $f_{D,k}$ is the Doppler shift, $\tau_k$ is the delay for target $k$, $T$ is the frame interval, $\Delta f$ is the subcarrier spacing, and $\tilde{\eta}$ is additive white Gaussian noise (AWGN).

Thus, the estimation of the round-trip delay and Doppler shift—and consequently, the determination of range and relative velocity of the targets, respectively —can be formulated as a spectral estimation problem. These parameters can be obtained for example via a two-dimensional Fourier transform (2D-DFT/FFT) as done in \cite{sanson2020ofdm}. The range resolution, $\Delta r$, is determined  by the total bandwidth $B$ of the transmitted signal and is given by  $\Delta r = \frac{c}{2B} = \frac{c}{2N \Delta f}$,  where $B = N \Delta f$. The Doppler resolution depends on the number of LTF frames $M$ and the time interval $T$ between frames (i.e., the frame rate), and is expressed as $\Delta v = \frac{c}{2M f_c T} = \frac{c \Delta f}{2M f_c}$

\subsection{Time-Phase Synchronization}
Commercial Wi-Fi systems are affected by HW asynchronization (delay and phase offsets). To address these, we apply the algorithm described in~\cite{sansonRange}, which performs delay and phase alignment. We first perform the delay calibration, by computing the cross-correlation between the known training-symbol sequence to perform the coarse synchronization, yielding $l_{\text{coarse}} = \arg\max_{l} C(l)$. Then, a refined correlation (upsampled by factor $U$) gives the fine delay $l_{\text{fine}} = \arg\max_{l} C_{\text{fine}}(l)$. The total delay correction is $l_{\text{eff}} = l_{\text{coarse}} + l_{\text{fine}}$ \cite{sansonRange}. 

In order to perform the phase synchronization, for each frame $m$, the average phase is computed as
\begin{equation}
\theta_m = \angle \left(\frac{1}{N} \sum_{n=1}^{N} D(m,n)\right)
\end{equation}
The reference phase $\phi_{m-1}$ is the average over the previous $H$ frames. The phase difference is $\Delta\theta_m = \phi_{m-1} - \theta_m$, quantized in steps of $\delta$, with
$\text{fix}_m = \operatorname{round}\left(\frac{\Delta \theta_m}{\delta}\right) \cdot \delta$. The phase of all subcarriers is then corrected by 
$D(m,n) \leftarrow D(m,n) \cdot \exp\left(j\,\text{fix}_m\right)$ \cite{sansonRange}.

\section{Range-Filtered Doppler Spectrum (RF-DS)}
\label{sec:rfds}

Unlike traditional approaches that compute a full RDM via 2D-FFT~\cite{sansonRange}, our method achieves range-selective Doppler estimation through: (1) finite impulse response (FIR) filter-based self-interference cancellation that provides fine-grained zero-Doppler control, and (2) efficient matched filtering in the frequency domain for range-specific Doppler spectra, generating a time/Doppler map. These techniques enable low-complexity, real-time operation on battery-powered devices.

\subsection{Enhanced Self-Interference Cancellation via FIR Filtering}
In monostatic  sensing, self-interference (SI) from Tx-Rx coupling and static environmental reflections dominate the received signal, appearing as strong zero-Doppler components that mask moving targets. In \cite{sansonRange} the authors use Direct Current (DC) removal by subtracting the mean across all $M$ frames $\hat{D}(m,n) = D(m,n) - \frac{1}{M} \sum_{m=0}^{M-1} D(m,n)$. The Doppler frequency band cancellation of this technique depends on the number of frames $M$ used in the Doppler FFT window. The DC removal does not provide good control over the rejected Doppler band and can suppress slow motions (e.g., breathing at 0.08 m/s). 

To address these limitation we employ Moving Target Indication (MTI) filters to remove stationary or slow-moving objects, such as those originating from wall reflections or Tx/Rx leakage. The filter used in this work is a high band-pass FIR filter with 64 steps, designed with cosine windows. The normalized Doppler frequency response $H(\omega_D T)$ of an FIR filter with normalized angular Doppler frequency $\omega_D T$ is computed as \cite{ash2018digital}:
\begin{equation}
H(\omega_D T) = \sum_{k=0}^{M} b_k e^{-j\omega_D T k}
\end{equation}
By providing finer control over zero-Doppler rejection, the FIR filter enables detection of subtle slow motions (e.g., user breathing while stationary at desk, velocity $\approx$5--10 mm/s) that would be masked by coarse DC removal.

To improve processing efficiency, the FIR filter is not applied directly to the raw CSI data. Instead, it is applied to each estimated range gate. This approach significantly reduces computational load, as the filter only processes the selected range gates—typically far fewer than the total number of subcarriers—rather than filtering across all subcarriers.  The FIR filter requires $M_{\text{fir}} \times R_g$(range gates)  complex multiplications and additions per frame.

\subsection{Range-Specific Doppler Spectrum via Matched Filtering}
Traditional RDM requires computing FFT/DFT across all $N$ subcarriers, then FFT across time for all $N$ range bins. For HPD, we only need Doppler information at $R_g$ specific ranges (e.g., $R_g=9$ gates).

We compute the Doppler spectrum at range  gate $R_i$ directly via phase-compensated summation across subcarriers:
\begin{equation}
s_i(m) = \sum_{n=0}^{N-1} D(m,n) \cdot e^{j \phi_n(R_i)} \cdot w(n)
\label{eq:single_point_dft}
\end{equation}
where $\phi_n(R_i) = -2\pi n \Delta f \cdot 2R_i / c$ is the phase shift for range $R_i$ at subcarrier $n$, and $w(n)$ is an optional window function.  This operation is equivalent to matched filtering in the frequency domain or single-point DFT. By multiplying each subcarrier by $e^{j \phi_n(R_i)}$, we \emph{phase-align} the contributions from range $R_i$ across all frequencies, causing them to add coherently when summed. Signals from other ranges add incoherently and are attenuated. The Doppler spectrum is then:
\begin{equation}
S(f_D, R_i) = \text{FFT}_m\{s_i(m)\}
\label{eq:doppler_fft}
\end{equation}

To track the temporal evolution of target motion, we accumulate Doppler spectra from a sliding window of $M_w$ successive processing windows. This forms a time–Doppler map for range gate $R_i$:
\begin{equation}
\mathcal{S}_i(t_w, f_D) = S_w(f_D, R_i), \quad w \in [w_{\text{current}} - M_w + 1,\, w_{\text{current}}]
\label{eq:time_doppler_map}
\end{equation}
where $t_w$ is the time index of window $w$. This 2D representation ($M_w \times$ Doppler bins) provides temporal context for presence detection decisions.

\subsection{Feature Extraction and Presence Detection}
\subsubsection{Signal to Noise Ratio (SNR) Estimation}
To enable robust target detection under varying environmental conditions, we estimate the noise floor adaptively for each range gate. Within the sliding window of $M_W$ frames, we first compute the mean Doppler power:
\begin{equation}
\mu_{\text{raw},i} = \frac{1}{M_W} \sum_{w=1}^{M_W} P_i(w)
\end{equation}
where $P_i(w)$ is the peak Doppler power in range gate $i$ at window index $w$. To mitigate contamination from strong transient targets, we apply clipping before computing the final noise floor estimate:
\begin{equation}
P_i^{\text{clip}}(w) = \min\left(P_i(w),\, \mu_{\text{raw},i} + \Delta P_{\text{clip}}\right)
\end{equation}
The clipped noise floor is then
\begin{equation}
\mu_{\text{noise},i} = \frac{1}{W} \sum_{w=1}^{W} P_i^{\text{clip}}(w)
\end{equation}
and the SNR for each range gate is computed in dB as $\text{SNR}_i = P_i - \mu_{\text{noise},i} \quad$.

\subsubsection{Presence Detection}
To perform user presence detection, we employ a detection time window approach, denoted as $W_{\text{det}}$, in which the Doppler bins are averaged over the window to estimate target presence. By computing the mean Doppler power within each $W_{\text{det}}$ cell before applying the SNR detection stage, the system achieves less false negatives enhancing sensitivity to subtle user activity, as shown in the results section. Following this, the system selects the range gate with the highest SNR:
\begin{equation}
i^* = \arg\max_i \text{SNR}_i
\end{equation}
Additionally, to increase the range accuracy, we perform quadratic interpolation on the range gate index based on the magnitude of the signal. To further improve robustness and reduce false triggers, we use a multi-frame majority voting scheme for the final HPD state decision.

For human presence detection on a laptop, the most relevant spatial region is immediately in front of the device where the user typically interacts. We define range gates corresponding to the following HPD states:
\begin{itemize}
\item \textbf{Near zone:} $R_1 = 0$--2 m (user at desk)
\item \textbf{Approach zone:} $R_2 = 2$--5 m (approaching/leaving)
\item \textbf{Far zone:} $R_3 > 5$ m (no presence)
\end{itemize}

\section{Experimental Setup}
\label{sec:experimental_setup}

To validate the proposed RF-DS method, we conducted a series of experiments assessing its ability to detect slow micro-motions (breathing), track user approach and leave events, and generalize across different Wi-Fi HW platforms.

\begin{figure}[b]
    \centering
    \begin{subfigure}[b]{0.48\linewidth}
        \includegraphics[width=\linewidth]{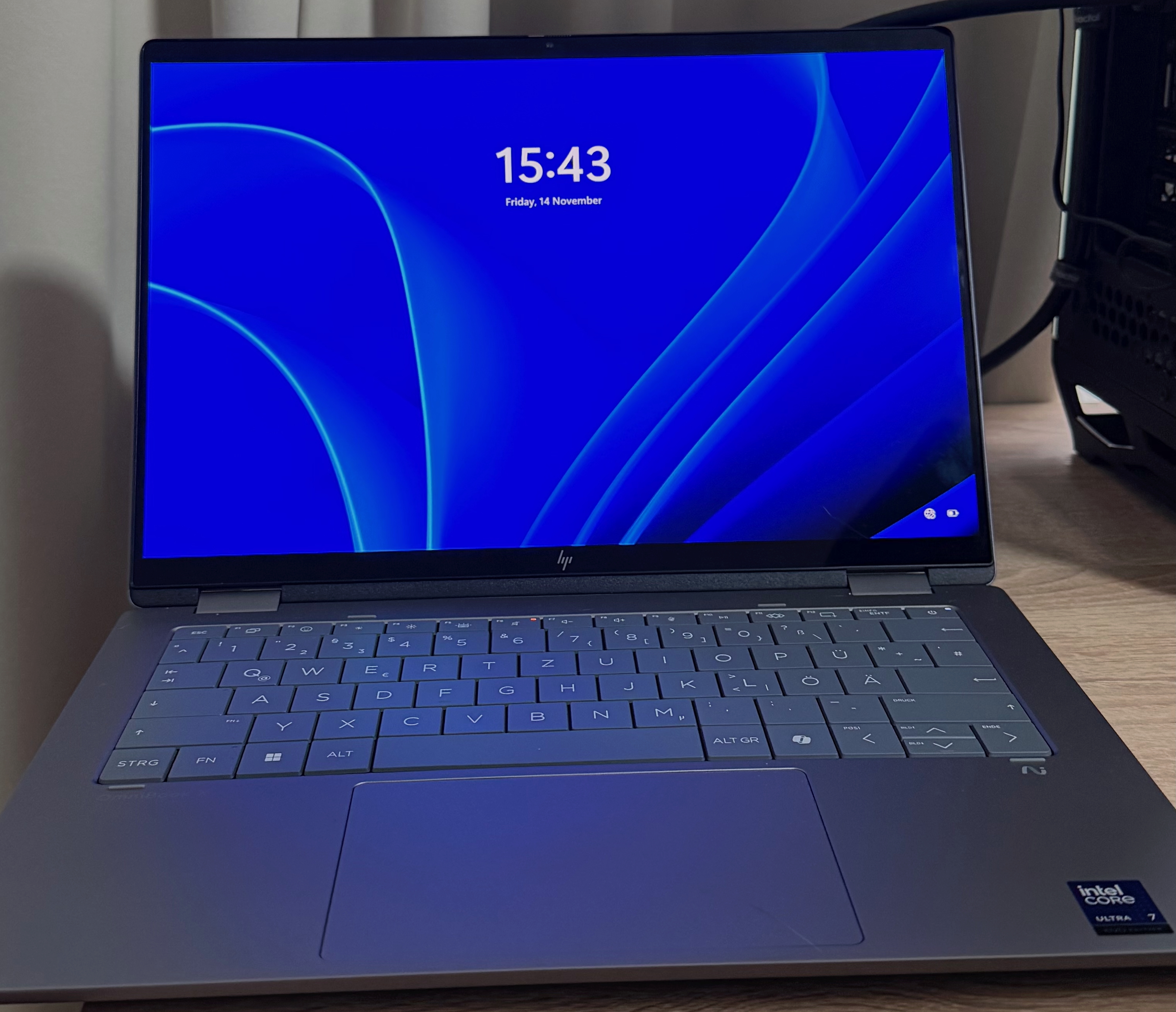}
        \caption{HP laptop (Wi-Fi 7)}
    \end{subfigure}
    \hfill
    \begin{subfigure}[b]{0.485\linewidth}
        \includegraphics[width=\linewidth]{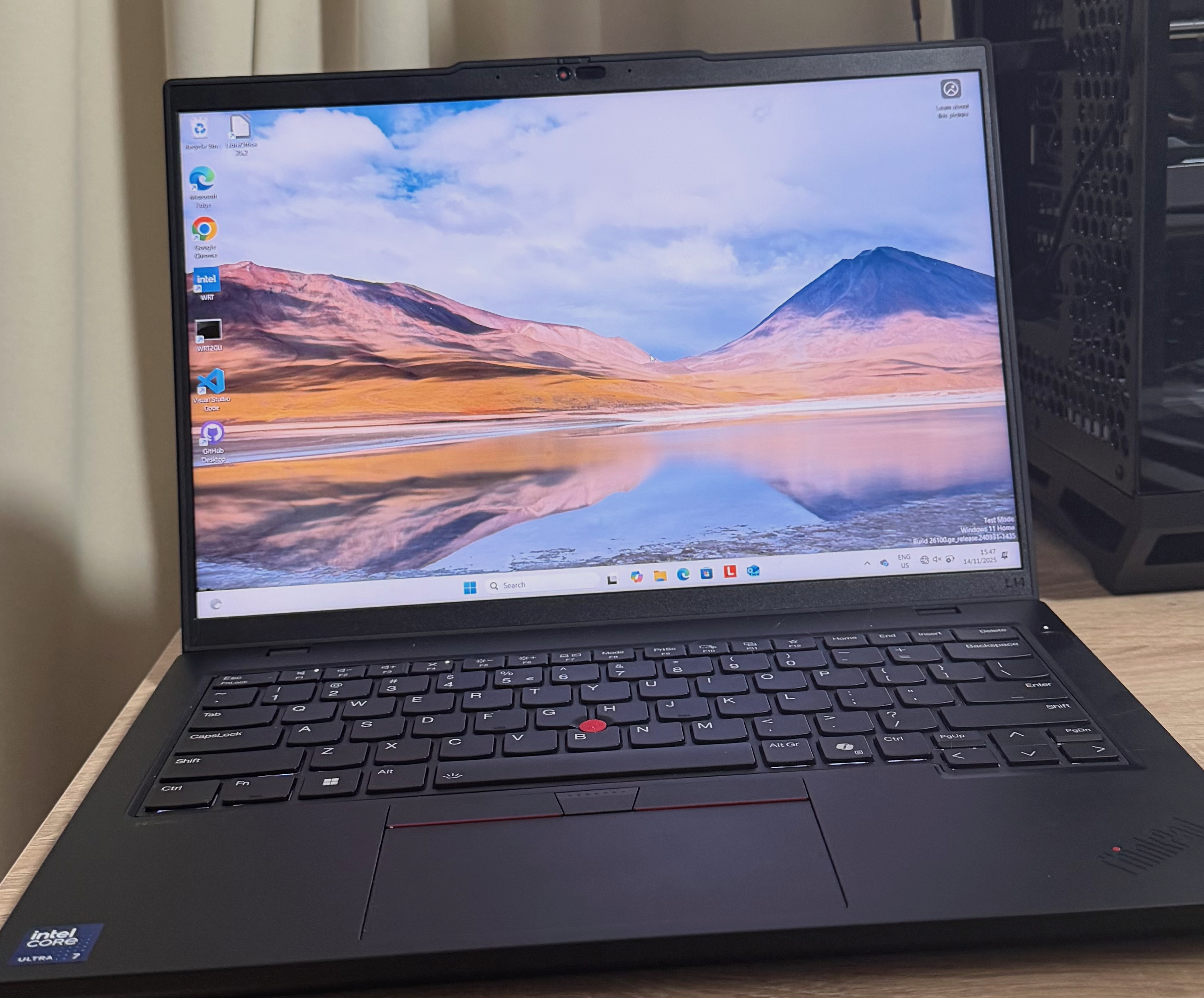}
        \caption{Lenovo laptop (Wi-Fi 6E)}
    \end{subfigure}
    \caption{Devices used for the measurements: (a) HP laptop (Wi-Fi 7), (b) Lenovo laptop (Wi-Fi 6E).}
    \label{fig:breathing_detection}
\end{figure}

For slow-motion detection, a user remained stationary and breathed normally at a distance of 3 m from the laptop. This scenario was used to evaluate whether the RF-DS algorithm could reliably isolate micro-Doppler signatures of breathing within the appropriate range gate. The approach–leave experiment involved a user moving from a seated position near the laptop (0.5 m) to a distance of 8 m, briefly pausing, and then returning to the starting position. To assess cross-platform robustness, the approach–leave protocol was repeated in a different device. All experiments were performed in a furnished standard office environment to ensure realistic multipath and clutter conditions. Ground truth for user position and motion was established via synchronized video recordings aligned with the CSI data capture.

\begin{table}[t]
\centering
\caption{RF-DS System Parameters for Idle and Detection Modes}
\label{tab:system_parameters}
\small
\begin{tabular}{lccccl}
\hline
\hline
\textbf{Parameter} & \textbf{Symbol} & \textbf{Unit} & \textbf{Idle} & \textbf{Detection} \\
\hline
\hline
Frame Rate                & $f_{\text{frame}}$   & Hz      & 10      & 100     \\
Number of Frames          & $M$                  & --      & 32      & 32      \\
Bandwidth                 & $B$                  & MHz     & 160     & 160     \\
Carrier Frequency         & $f_c$                & GHz     & 5.8     & 5.8     \\
Range Resolution          & $\Delta r$           & m       & 0.94    & 0.94    \\
Number of Range Gates     & $R_g$                & --      & 9       & 9       \\
Velocity Resolution       & $\Delta v$           & m/s     & 0.004   & 0.074   \\
Unambiguous Velocity      & $v_{\max}$           & m/s     & $\pm$0.12 & $\pm$1.19 \\
SNR Window                & $M_W$                & frames  & 20      & 20      \\
Detection Window          & $W_{\text{det}}$     & frames  & 3       & 3       \\
Detection Threshold       & $T_{\text{SNR}}$     & dB      & 12      & 12      \\
Presence State Window     & $M_{\text{maj}}$     & frames  & 3       & 3       \\
\hline
\end{tabular}
\end{table}
\textbf{HW Platform and System Parameters:}
The primary experiments were conducted using an HP laptop with Wi-Fi 7, while cross-platform tests included a Lenovo ThinkPad with Wi-Fi 6E. CSI data was acquired using a modified Intel driver, capable of capturing LTF symbols from each frame. Both platforms operated in a monostatic sensing configuration, employing a single Tx and Rx antenna. The key system parameters for both \textit{Idle} (low-rate) and \textit{Detection} (high-rate) operation modes are summarized in Table~\ref{tab:system_parameters}. These settings were maintained consistently across all HW platforms and experiments. For spectral SNR estimation within each range gate, a temporal window was used. The detection threshold was 12 dB, and presence state estimation was determined using a 3-frame majority label rule. The system switched to \textit{Idle} mode after 10 negative presence estimations, and switched back to \textit{Detection} mode after one detection.

\begin{figure}[b]
    \centering
    \includegraphics[width=0.485\linewidth]{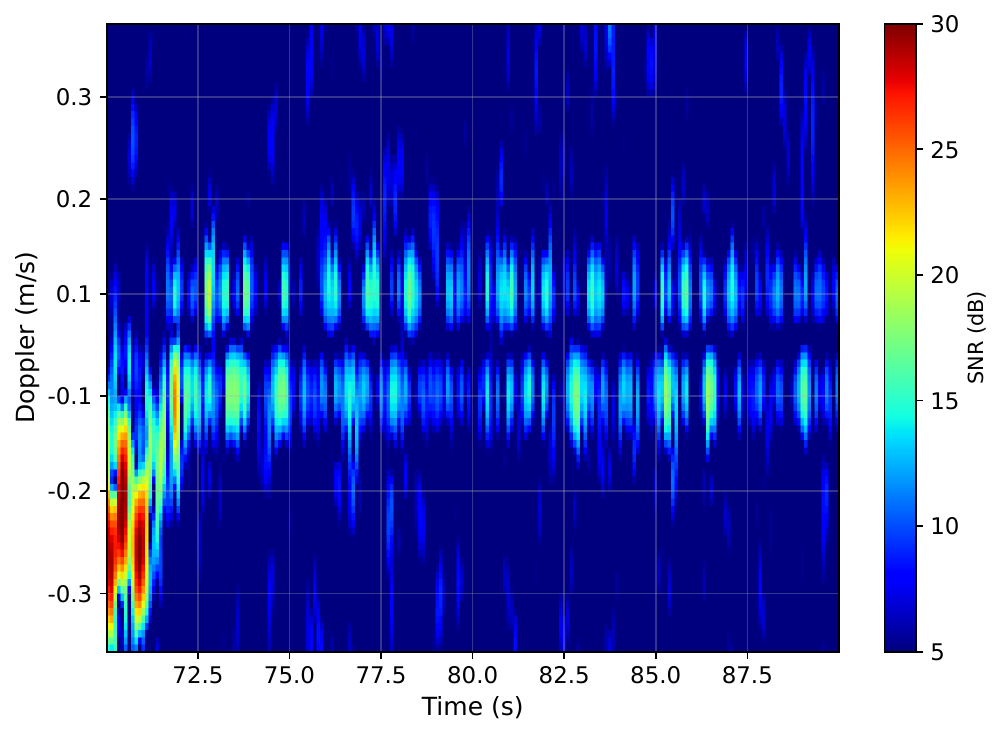}
    \includegraphics[width=0.485\linewidth]{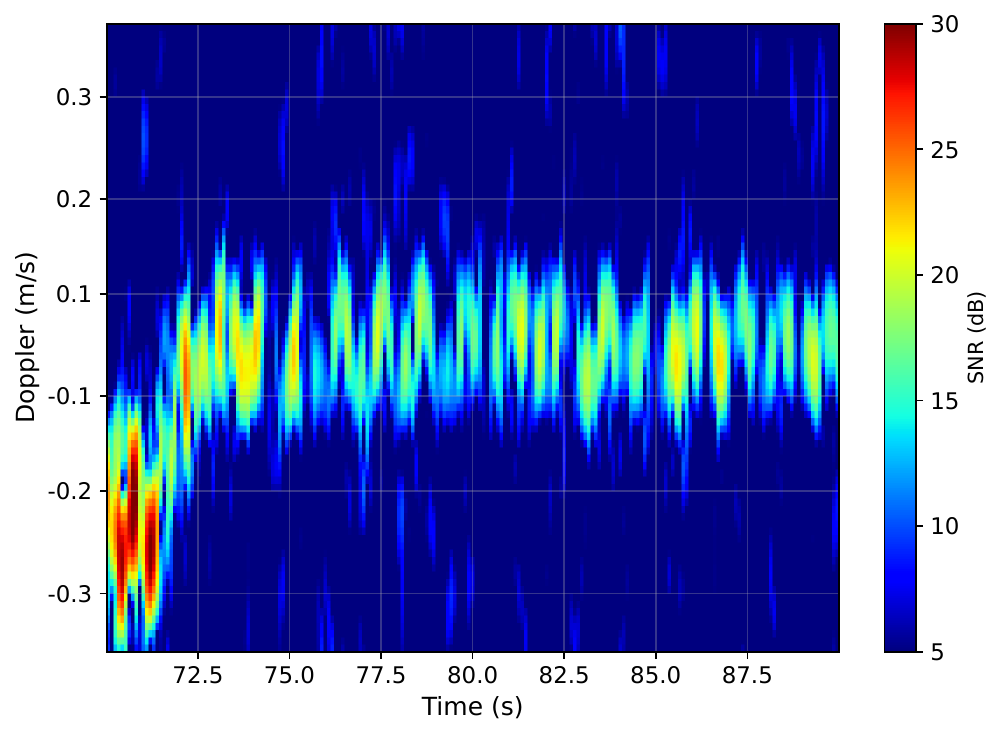}
    \caption{RF-DS Doppler spectra for a single range (3 m) with user breathing
    (a) using DC cancellation \cite{sansonRange} (b) proposed MTI filtering. 
    MTI filtering improves self-interference suppression and enhances micro-Doppler visibility of respiration.}
    \label{fig:breathing_detection}
\end{figure}

\begin{figure*}[h]
    \centering
    \begin{subfigure}[b]{0.48\linewidth}
        \includegraphics[width=\linewidth]{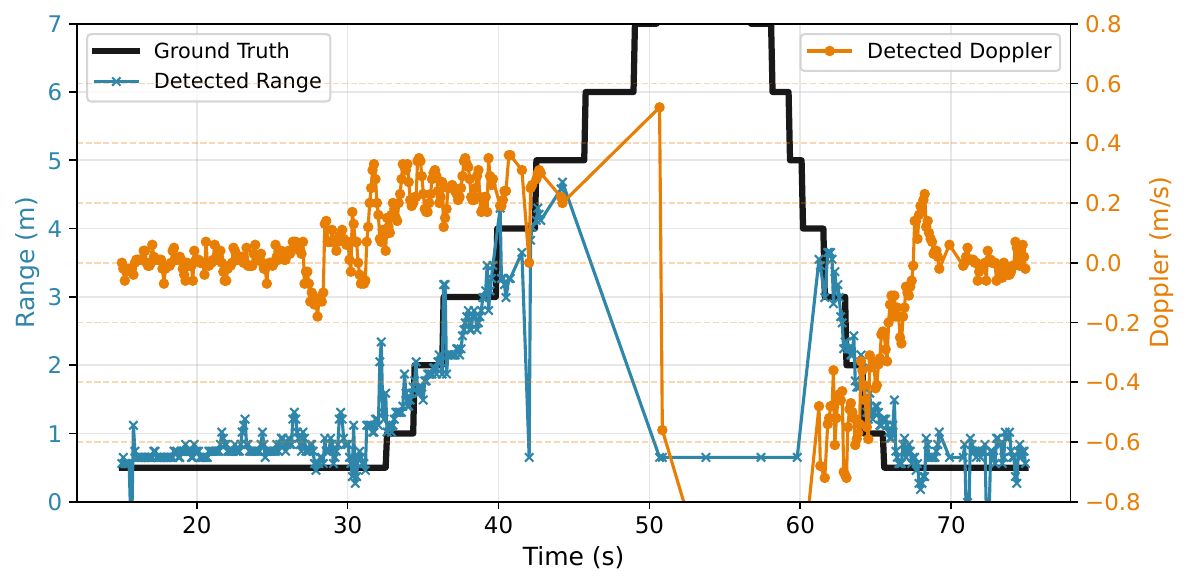}
        \caption{2D FFT range--Doppler map}
    \end{subfigure}
    \hfill
    \begin{subfigure}[b]{0.49\linewidth}
        \includegraphics[width=\linewidth]{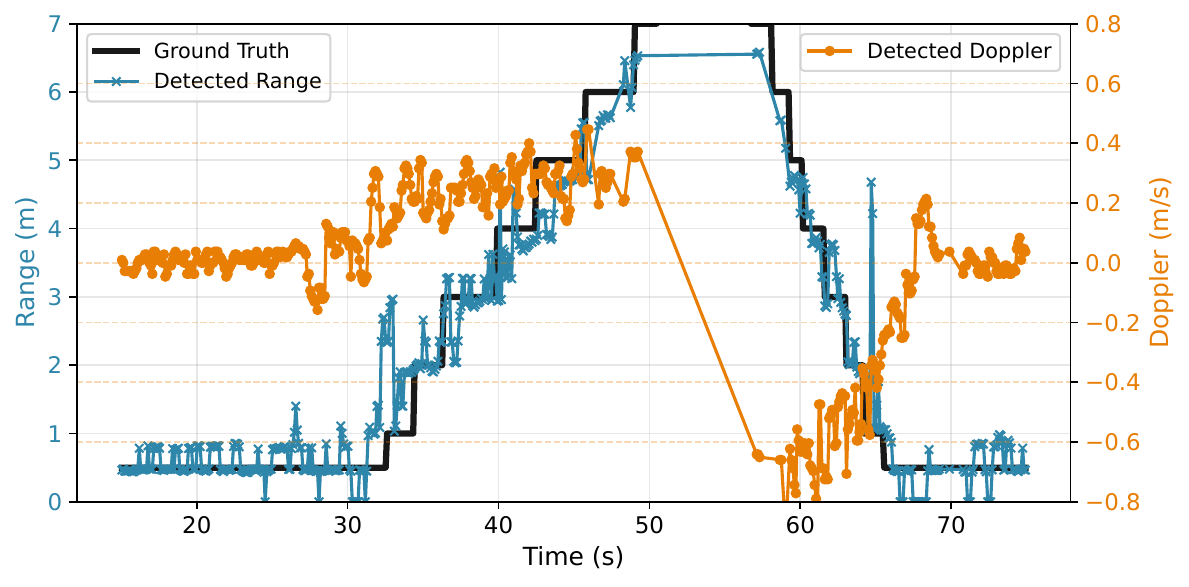}
        \caption{RF-DS output}
    \end{subfigure}
    \caption{Range and velocity tracking during the approach--leave cycle. (a) 2D FFT range--Doppler map; (b) RF-DS output. Left axis: Estimated range (blue) vs. ground truth (black); right axis: Doppler velocity (red). RF-DS yields smoother, more robust tracking for user presence applications.}
    \label{fig:approach_leave_results}
\end{figure*}

The classification of the HPD state was based on both SNR thresholds and estimated range $\hat{r}_{i}$ as follows:

\textbf{Approaching/leaving}: $\text{SNR}_{i}>10$dB and $2 \text{m}<|\hat{r}_{i}|<5\text{m}$

\textbf{Presence}: $\text{SNR}_{i} > 10$ dB and $|\hat{r}_{i}| < 2 \text{m}$

\textbf{No presence}: $\text{SNR}_{i} < 10$ dB or $|\hat{r}_{i}| > 5 \text{m}$
\

\section{Results and Evaluation}
\label{sec:results}

\subsection{Breathing Detection}

The first evaluation tests the ability of RF‑DS to detect user breathing at a distance of 3m, a challenging range due to the very small Doppler shifts produced by respiratory motion. We compare two SI‑mitigation approaches—traditional DC cancellation \cite{sansonRange} and the proposed MTI filter—while keeping the RF‑DS processing identical in both cases. As shown in Fig.~\ref{fig:breathing_detection}, the MTI filter yields a noticeably cleaner and more pronounced micro‑Doppler response, improving sensitivity to the low‑frequency breathing components. This enhancement is crucial for presence‑detection scenarios in which a weak breathing signal due to user position may need to be detected. The results confirm that effective SI mitigation is key to enabling reliable micro‑motion detection at extended ranges.

\subsection{Approach-Leave Detection }

Figure~\ref{fig:approach_leave_results} presents representative results from an approach–leave cycle, illustrating the ability of the system to continuously track user movement between the laptop and a distance of over 6 m. In this test, the user began seated 0.5 m from the laptop, then after 30 sec stood up, walked away beyond 6 m, paused, and finally returned to the seat. Both the proposed RF-DS approach and a conventional 2D range–Doppler map were evaluated, with MTI filtering applied in both cases to enhance low-motion sensitivity.

The RF-DS method delivered more stable range and velocity estimates throughout the cycle, particularly during long-range and low-speed segments. Although RF-DS sacrificed some fine-grained range resolution compared to full 2D range–Doppler mapping, it provided smoother and more reliable detection in the relevant workspace zone, which is more critical than cm-level accuracy for user presence applications. Notably, RF-DS maintained robust detection at extended ranges, supporting features such as automated sleep and wake control for the laptop. This stability and sensitivity are attributable to the temporal windowing and spectral estimation inherent in the RF-DS approach, which enables more consistent SNR and target detection than single-frame range–Doppler maps.


\begin{figure}[b!]
    \centering
    \begin{subfigure}[b]{0.99\linewidth}
        \includegraphics[width=\linewidth]{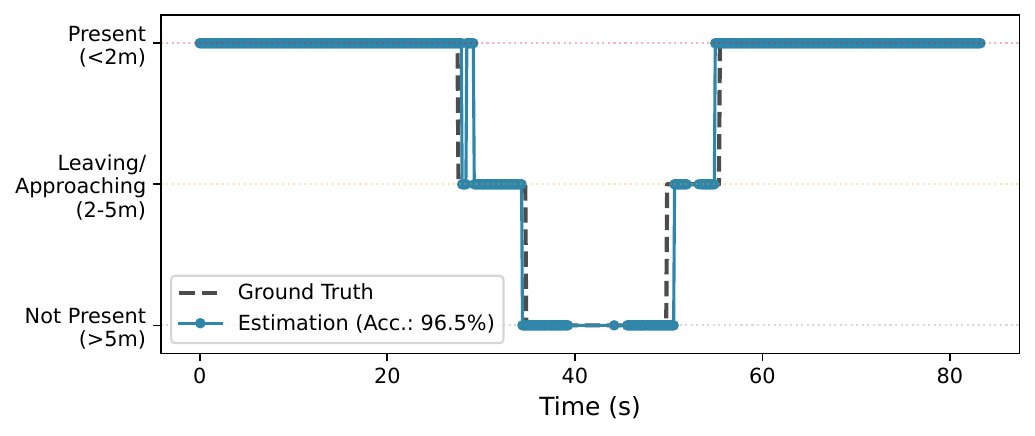}
        \caption{Lenovo Wi-Fi 6E}
    \end{subfigure}
    \\
    \begin{subfigure}[b]{0.99\linewidth}
        \includegraphics[width=\linewidth]{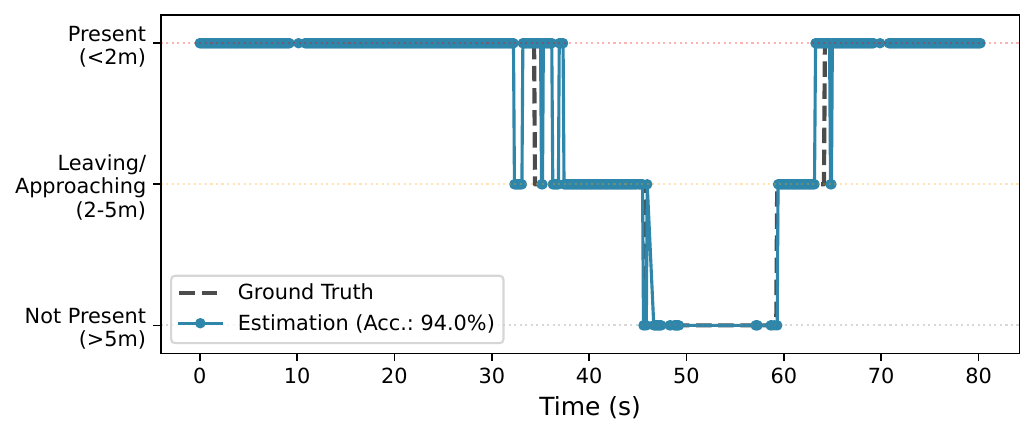}
        \caption{HP Wi-Fi 7}
    \end{subfigure}
    \caption{Cross-platform HPD results: Presence, absence, approach, and leave state estimation for (a) Lenovo Wi-Fi 6E and (b) HP Wi-Fi 7 laptops. Both platforms demonstrate consistent performance, confirming the HW independence of the proposed approach.}
    \label{fig:proximity_comparison_target}
\end{figure}

\textbf{HPD and Cross-Platform Consistency:} 
Figure~\ref{fig:proximity_comparison_target} illustrates the performance of the proposed Wi-Fi-based HPD technology in detecting user presence, approach, leave, and absence states on two commercial laptop platforms: a Lenovo with Wi-Fi 6E and an HP with Wi-Fi 7. The system achieved similar accuracy on both platforms (94\% and 96.5\%, respectively), indicating strong HW generalizability and confirming the suitability of this approach for laptop HPD applications regardless of the Wi-Fi NIC generation used.

The state estimation latency—measured from the initiation of user movement (approaching or leaving) to state classification—averaged 0.48 sec (maximum latency was 0.89 sec), which is sufficient for the laptop to trigger wake operations before the user arrives. The figures further demonstrate reliable classification of presence, absence, and transitional states, even without any calibration, environmental adaptation, or HW modifications.

These results underscore the practicality of leveraging existing Wi-Fi NICs in commercial laptops for HPD, enabling robust presence sensing without specialized HW or prior environment measurements. Future improvements will focus on refining motion classification to distinguish between humans, animals, and inanimate objects, as preliminary findings suggest these sources exhibit distinct Doppler profiles that can be exploited to further reduce false positives.

\section{Conclusion}
\label{sec:conclusion}

In this paper, we demonstrate for the first time a HPD solution based on Wi-Fi sensing that only requires the built-in Wi-Fi HW on commercial laptops, while requiring no external infrastructure such as access points. We introduce the RF-DS method, which enables robust detection of user presence, approach, and departure events while greatly reducing computational complexity compared to traditional full range–Doppler map techniques. The proposed system further incorporates adaptive multi-rate processing to minimize power consumption and is compatible with stringent operating system requirements for latency and energy efficiency.

Extensive experiments demonstrated that our method reliably detects both larger-scale user movements and subtle micro-motions such as breathing at a range of several meters. The RF-DS approach delivered stable and accurate state estimation across a range of real-world office environments, with no need for calibration or HW modification. It also generalized effectively across different commercial Wi-Fi NICs (Wi-Fi 6E and Wi-Fi 7). Achieving over 94\% accuracy in cross-platform tests and sub-second response latency, the system is well suited for practical HPD deployment, supporting seamless power management and privacy features in consumer laptops.

\bibliographystyle{./IEEEtran}
\bibliography{./IEEEabrv,./references}

\end{document}